# THE LATTICE BOLTZMANN EQUATION METHOD FOR THE SIMULATION OF COMPRESSIBLE FLUID FLOW

Yu Chen, Hirotada Ohashi, and Mamoru Akiyama
Department of Nuclear Engineering
The University of Tokyo
7-3-1 Hongo, Bunkyo-ku, Tokyo 113, Japan

## ABSTRACT

We systematically derived hydrodynamic equations and transport coefficients for a class of multi-speed lattice Boltzmann models in D dimensions, using the multi-scale technique. The constitutive relation of physical fluid is recovered by a modified equilibrium distribution in Maxwell-Boltzmann type. With the use of the rest particles and the particle reservoir, we were able to add one degree of freedom into the sound speed of the modeled fluid. When the sound speed is tuned small enough, the compressible region of fluid flow can be reached. An example 2-D model is presented, together with the numerical verification for its transport coefficients.

## 1 INTRODUCTION

As it is well known, the lattice Boltzmann models for fluid dynamics are extremely suitable for the modern super-computation. The realization is due to some fundamental characteristics of the discrete microscopic model: the intrinsic stability, the easiness in dealing with complicated geometric boundaries and the fast and local operations fit for massively parallel processing.

However, the conventional lattice Boltzmann model(McNamara and Zanetti, 1988) also suffers from some drawbacks. Apart from the doubt about the efficiency of the model on the serial computer frame, the main problems result from the unphysical terms appearing in the macro-dynamic equations. Specifically, the modeled fluid lacks of Galilean invariance(presence of a so-called $g(\rho)$ density dependent factor in the non-linear advection term) and contains a velocity dependent part in the pressure term.

These difficulties restrict the range within which the model may be applied to the simulation of fluid flow. Accordingly, the incompressible limit is caused by two factors. First, the lost Galilean invariance has to be recovered by rescaling the time with the coefficient $g(\rho)$. This manipulation requires the density of the modeled fluid to be set to a constant $\rho_0$. Secondly, the sound speed of the modeled fluid is generally determined by the lattice geometry in the microscopic world. As the velocity vectors of particles are identical to the unit base vectors of the regular polygon which tesselates the lattice space, the sound speed of the model turns out to be order 1. Whereas the equilibrium particle distribution function must be expanded in the small limit of flow velocity, that is $u \ll c_s$ in the hydrodynamic derivation of the LBE model. The Mach number thus has to be kept small, which means that the region where the compressibility becomes important cannot be reached easily.

Recently a lattice BGK model was proposed (Qian *et al.* , 1992) to recover the full Navier-Stokes equations for the isothermal fluid. The name comes from the adoption of the collision operator which had been employed in the kinetic BGK equation(Bhatnagar *et al.*, 1954). The operator takes a relaxation form in which the particle distribution approaches to the equilibrium value in a rate of $1/\tau$. Furthermore, all those unphysical terms were eliminated by employing the Maxwell-Boltzmann type of equilibrium distribution function. Unfortunately the sound speed of the model still remained frozen so that it was difficult to increase the Mach number. It was claimed that Mach number must be less than $\sqrt{5}/2$, through a stability analysis on the positivity requirement of the particle density. More severe conditions would be expected if this analysis were done on the spatial gradient of hydrodynamic quantities. Therefore some modification of the model is needed to solve the difficulty.

In this paper, we employ a particle reservoir to ensure the mass conservation while the unphysical terms are removed by a modified Maxwell-Boltzmann type of equilibrium distribution. With such operations, we set freely certain member of the $t_p$'s, which are factors of the density allocation on different sublattices. For example, in a 2-D 3-speed model $t_1$ and $t_2$(allocating factors for speed 1 and 2 particles) are designed to recover the isotropy for the fourth order tensor, while $t_0$(allocating factor for rest particles) is specified as the controller of sound speed of the modeled fluid.

We derive the hydrodynamic equations and the transport coefficients for a general D-dimensional model in Section 2. An example of 2-D model which consists of 9 types of particles will be introduced in Section 3. Results of numerical calculation of this model are presented in Section 4. The conclusion and perspective of the model will be discussed in Section 5.

## 2 LATTICE HYDRODYNAMICS

### 2.1 Lattice Field And Symmetric Properties

The lattice field defined in this model is composed of several sub-lattices whose base vectors are permutations of $(\pm 1, \ldots, \pm 1, 0, \ldots, 0)$ in $D$ dimensions. We use $p$, namely, the number of non-zero components in the base vector, as the index for different sub-lattices. The number of base vectors of each sub-lattice is denoted by $b_p$ and can be calculated by

$$b_p = \frac{2^p D}{p!(D-p)!} \,. \tag{1}$$

$c_{pi\alpha}(0 \le p \le D, 1 \le i \le b_p, 1 \le \alpha \le D)$ is used to denote the $\alpha$ component of possible velocity vectors for particles in cell $i$ of sub-lattice $p$. As $\boldsymbol{c}_{pi}$'s are identical to the base vectors, the following relationship should hold,

$$\sum_\alpha c_{pi\alpha}^2 = c_p^2 = p \,. \tag{2}$$

Furthermore, the velocity moments can be explicitly given by using the symmetric properties of the regular lattice. We have,

**VMP 1** $\sum_i c_{pi\alpha} = 0$.

**VMP 2** $\sum_i c_{pi\alpha} c_{pi\beta} = \frac{b_p c_p^2}{D} \delta_{\alpha\beta}$.

**VMP 3** $\sum_i c_{pi\alpha} c_{pi\beta} c_{pi\delta} = 0$.

**VMP 4** $\sum_i c_{pi\alpha} c_{pi\beta} c_{pi\delta} c_{pi\gamma} = \psi_p \gamma_{\alpha\beta\gamma\delta} + \varphi_p(\delta_{\alpha\beta}\delta_{\gamma\delta} + \delta_{\alpha\gamma}\delta_{\beta\delta} + \delta_{\alpha\delta}\delta_{\beta\gamma})$.

Note that the fourth velocity moment will be denoted as $T_{p\alpha\beta\gamma\delta}$ in this paper and $\psi_p$ and $\varphi_p$ are calculated by(Qian, 1990)

$$\psi_p = \frac{(D+2-3p) b_p c_p^4}{pD(D-1)} \,, \qquad \varphi_p = \frac{(p-1) b_p c_p^4}{pD(D-1)} \,. \tag{3}$$

These expressions can easily be verified by the value of $T_{p1111}$ and $T_{p1212}$. We define $\psi_p$ and $\varphi_p$ for rest particles as

$$\psi_0 = 0 \,, \qquad \varphi_0 = 0 \,. \tag{4}$$

### 2.2 Definition Of Macroscopic Quantities

The particle distribution in cell $i$ of sub-lattice $p$ at time t and lattice site $\boldsymbol{x}$ is designated as $N_{pi}(\boldsymbol{x}, t)$. At this stage we introduce a particle reservoir, by which the mass conservation may be ensured in the microscopic process(Chen *et al.*, 1992). The reservoir works in background at every lattice site and delivers products at every time step. The products, written as $R(\boldsymbol{x}, t)$ can be either positive or negative. When $R > 0$, they are taken as normal particles which may compensate the mass. Otherwise, $R$ is called as "particle-killer" which diminish the surplus of particles at that site. Hence the macroscopic density and momentum flux are defined as,

**DEF 1** $\rho(\boldsymbol{x}, t) = \sum_{p,i} N_{pi}(\boldsymbol{x}, t) + R$.

**DEF 2** $\rho(\boldsymbol{x}, t)\boldsymbol{u}(\boldsymbol{x}, t) = \sum_{p,i} N_{pi}(\boldsymbol{x}, t)\boldsymbol{c}_{pi}$.

The spatially uniform equilibrium can be reached in the lattice field when the macroscopic flow vanishes. In this case, the averaged particle density $d_p$ at each cell of the $p$ sub-lattice satisfies,

$$\sum_p b_p d_p = \rho \,. \tag{5}$$

An allocating factor, which is defined as $d_p = t_p \rho$, is used to rewrite equation (5) as

$$\sum_p b_p t_p = 1 \,. \tag{6}$$

Obviously $b_p t_p$ decides the percentage by which the particle density on the $p$ sub-lattice contributes to the macroscopic mass. We shall see later that $t_p$'s play important roles on recovery of isotropy of the fourth order tensor and lead into an adjustable sound speed for the modeled fluid. In order to ensure the isotropy of $T_{p\alpha\beta\gamma\delta}$, we must have

$$\sum_p t_p \psi_p = 0 \,. \tag{7}$$

We may simply take this requirement as an assumption at the present moment.

### 2.3 Equilibrium Distribution

We modify the extended Maxwell-Boltzmann type of equilibrium distribution in the following way so that both the Galilean invariance and the physical pressure term could be achieved in the sequel derivations. We write, when $u \ll 1$,

$$N_{pi}^{(0)} = m_0 + d_p[1 + \beta_0(c_{pi\alpha}u_\alpha) + h_1 u^2 + \frac{1}{2}\beta_0'(c_{pi\alpha}u_\alpha)^2] \,, \tag{8}$$

for moving particles,

$$N_{0i}^{(0)} = d_0 \,, \tag{9}$$

for rest particles and

$$R = \gamma u^2 - \sum_{p=1}^{D} b_p m_0 \,, \tag{10}$$

for the products of the particle reservoir. The $\beta_0, h_1, \beta_0'$ and $\gamma$ appearing above are given as follows

$$\beta_0 = \frac{D}{\sum_p t_p b_p c_p^2} \,, \tag{11}$$

$$h_1 = -\frac{1}{2}\beta_0 \,, \tag{12}$$

$$\beta_0' = \frac{1}{\sum_p t_p \varphi_p} \,, \tag{13}$$

$$\gamma = -\rho \left[ \frac{1 - \beta_0^2(1 - t_0 b_0)\sum_p t_p \varphi_p}{2\beta_0 \sum_p t_p \varphi_p} \right] \,. \tag{14}$$

These expressions are derived with the use of **VMP 1** $\sim$ **VMP 4** and **DEF 1** $\sim$ **DEF 2**. $m_0$ is an arbitrary constant, introduced to avoid the particle distribution to be negative.

### 2.4 LBE And Conservation Laws

The lattice Boltzmann equation is expressed in the ensemble averaged form of the micro-dynamic equation for particles moving on sub-lattices. It is given as follows,

$$N_{pi}(\boldsymbol{x}+\boldsymbol{c}_{pi}, t+1) - N_{pi}(\boldsymbol{x}, t) = \Delta_{pi}(N)\,, \qquad (15)$$

with the Boltzmann approximation applied to the collision term $\Delta_{pi}$. Thus the LBE exactly describes the two-step motion of particles in the micro-world. The first step is the streaming period during which the local particle distribution is transported to one of its nearest neighbors. The second one is the collision period during which the particle distribution relaxes to its equilibrium value which locally depends on the macroscopic density and momentum. $\Delta_{pi}$ may be linearized in the form

$$\Delta_{pi}(N) = \Delta_{pi}(N^{(0)}) + \Delta^{pi}_{qj}(N_{qj} - N^{(0)}_{qj})\,. \qquad (16)$$

From the definition of the collision process, we conclude that the first term will vanish as $N^{(0)}_{pi}$ is already the equilibrium value. $\Delta^{pi}_{qj}$ is the so-called collision operator which gives the probability for particle transition from the cell $j$ of sub-lattice $q$ to the cell $i$ of sub-lattice $p$. The operator of BGK type takes the simplest but most efficient form as

$$\Delta^{pi}_{qj} = -\frac{1}{\tau}\delta_{pq}\delta_{ij}\,. \qquad (17)$$

The mass and momentum are conserved globally in the streaming period while they are kept constant locally in the collision process. Such conservation relation may be described as follows,

$$\sum_{p,i}\Delta_{pi}(N) = 0\,, \sum_{p,i}\Delta_{pi}(N)\boldsymbol{c}_{pi} = 0\,. \qquad (18)$$

These relations would lead to the macroscopic conservation laws by substituting the equilibrium distribution function and rescaling time and space properly.

### 2.5 Multi-Scale Technique

In the long wavelength and low frequency limit, the LBE can be expanded as follows,

$$\partial_t N_{pi} + c_{pi\alpha}\partial_\alpha N_{pi} + \frac{1}{2}\partial_t^2 N_{pi} + c_{pi\alpha}\partial_t\partial_\alpha N_{pi} + \frac{1}{2}c_{pi\alpha}c_{pi\beta}\partial_\alpha\partial_\beta N_{pi} = \Delta_{pi}(N)\,. \qquad (19)$$

The multi-scale technique is applied in the following way. First we find a proper quantity $\epsilon \ll 1$ and write $N_{pi}$ perturbatively with the use of $\epsilon$,

$$N_{pi} = N^{(0)}_{pi} + \epsilon N^{(1)}_{pi}\,. \qquad (20)$$

Note that the perturbative part must satisfy the following constraints,

$$\sum_{p,i} N^{(1)}_{pi} = 0\,, \qquad \sum_{p,i} N^{(1)}_{pi}\boldsymbol{c}_{pi} = 0\,, \qquad (21)$$

due to the definition of the mass and momentum and the conservation relation (18). The time and space differential operator will also be replaced by

$$\partial_t \mapsto (\epsilon\partial_{t1} + \epsilon^2\partial_{t2})\,, \partial_\alpha \mapsto \epsilon\partial_\alpha\,. \qquad (22)$$

When the expressions stated above are substituted into equation (19), we shall pick up terms on the same order and formalize them in different equations.

#### 2.5.1 On $\epsilon$ Order

As terms on this order are considered, we shall have

$$\partial_{t1}N^{(0)}_{pi} + c_{pi\alpha}\partial_\alpha N^{(0)}_{pi} = -\frac{1}{\tau}N^{(1)}_{pi}\,. \qquad (23)$$

Next the equation (23) and its velocity moment will be integrated in the discrete velocity space. Employing **VMP 1 ~ VMP 4**, **DEF 1 ~ DEF 2** and equation (18), (21), two macroscopic equations appear below,

$$\partial_{t1}\rho + \partial_\beta(\rho u_\beta) = 0\,, \qquad (24)$$

$$\partial_{t1}(\rho u_\alpha) + \partial_\beta(\rho u_\alpha u_\beta) = -\frac{1}{\beta_0}\partial_\alpha\rho\,. \qquad (25)$$

#### 2.5.2 On $\epsilon^2$ Order

The terms kept on this order are formalized as follows,

$$\partial_{t1}N^{(1)}_{pi} + c_{pi\alpha}\partial_\alpha N^{(1)}_{pi} + \partial_{t2}N^{(0)}_{pi} + \frac{1}{2}\partial_{t1}^2 N^{(0)}_{pi} + c_{pi\alpha}\partial_{t1}\partial_\alpha N^{(0)}_{pi} + \frac{1}{2}c_{pi\alpha}c_{pi\beta}\partial_\alpha\partial_\beta N^{(0)}_{pi} = 0\,. \qquad (26)$$

The integration would be carried out in the same way as the first order equation. The integrated terms must be identified with the relations mentioned in the previous section, together with equation (24) and (25) themselves. Furthermore, the expression for $N^{(1)}_{pi}$ can be found from equation (23) and the order would be kept only to $\epsilon^2 u$ in the reduced equations. They are described below,

$$\partial_{t2}\rho = 0\,, \qquad (27)$$

$$\partial_{t2}(\rho u_\alpha) + (\tau - \frac{1}{2})\frac{1}{\beta_0}\partial_\alpha[\partial_\gamma(\rho u_\gamma)] - (\tau - \frac{1}{2})\sum_p t_p\varphi_p \beta_0\partial_\beta[\partial_\gamma(\rho u_\gamma)\delta_{\alpha\beta} + \partial_\alpha(\rho u_\beta) + \partial_\beta(\rho u_\alpha)] = 0\,. \qquad (28)$$

### 2.6 Macroscopic Equations

When equations (24), (25) and (27), (28) are recombined by using the relation (22), the full Navier-Stokes equations emerge as follows,

$$\partial_t\rho + \partial_\beta(\rho u_\beta) = 0\,, \qquad (29)$$

$$\partial_t(\rho u_\alpha) + \partial_\beta(\rho u_\alpha u_\beta) = -c_s^2\partial_\alpha\rho + \eta\partial_\beta^2(\rho u_\alpha) + \zeta\partial_\alpha[\partial_\gamma(\rho u_\gamma)]\,. \qquad (30)$$

the transport coefficients are given by

$$c_s = \sqrt{\frac{1}{\beta_0}}\,, \qquad (31)$$

$$\eta = (\tau - \frac{1}{2})\beta_0\sum_p t_p\varphi_p\,, \qquad (32)$$

$$\zeta = (\tau - \frac{1}{2})(2\beta_0\sum_p t_p\varphi_p - \frac{1}{\beta_0})\,. \qquad (33)$$

The derivation of hydrodynamic equations is completed at this point. However, the allocating factors($t_p$'s) are still to be determined by relation (6) and (7) so that the transport coefficients can be calculated explicitly. In the next section we shall give a 2 dimensional model and show $\tau$ and $t_p$'s influence on the viscosity and sound speed of the modeled fluid.

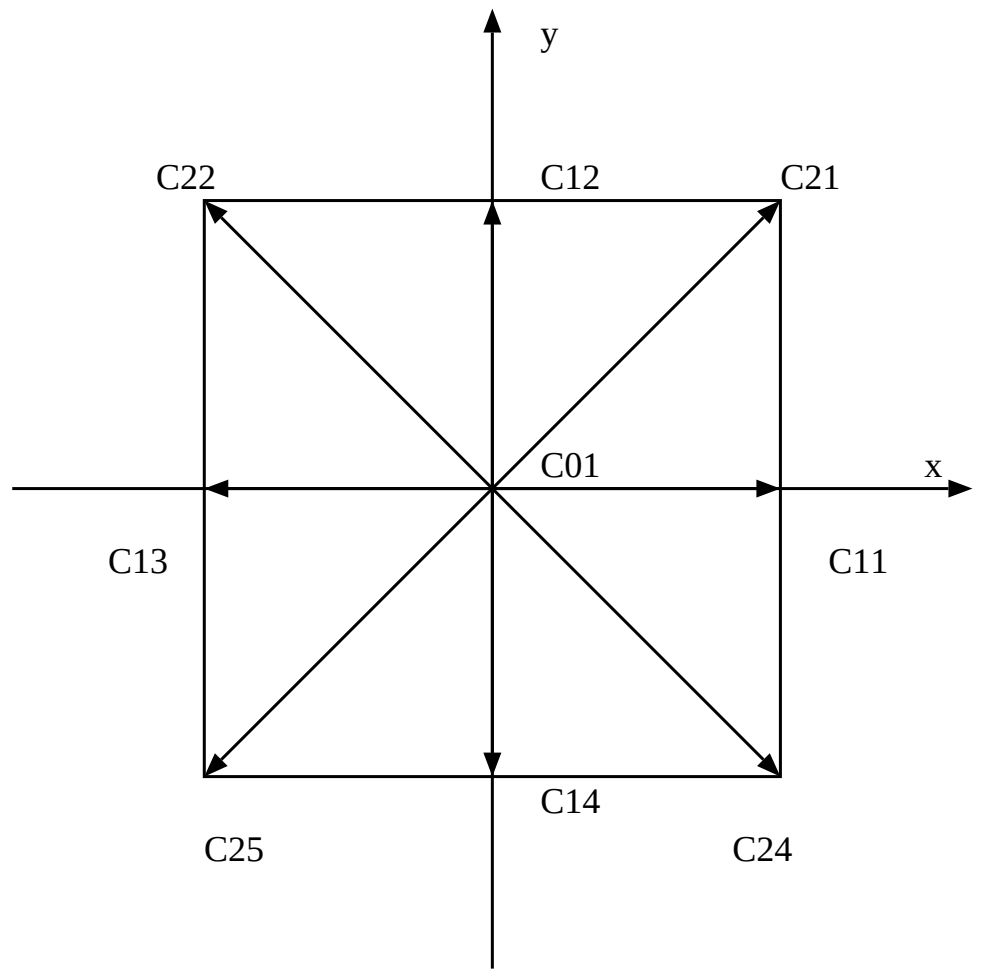

Figure 1: Lattice geometry for a 3-speed model in two dimensions. The arrows represent the velocity vectors of different types of particles.

# 3 AN EXAMPLE 2-D MODEL

We present a 2-D model which at most contains 9 kinds of particles at one lattice site. The lattice employed in the model is sketched in figure 1. We may immediately obtain the set of parameters concerning with the lattice geometry. They are classified by the lattice index $p$, whose possible values would be $0, 1, 2$ in this case,

$$\begin{cases} b_0 = 1 \\ b_1 = 4 \\ b_2 = 4 \end{cases}, \begin{cases} \psi_0 = 0 \\ \psi_1 = 2 \\ \psi_2 = -8 \end{cases}, \begin{cases} \varphi_0 = 0 \\ \varphi_1 = 0 \\ \varphi_2 = 4 \end{cases}. \quad (34)$$

In order to calculate the equilibrium distribution and the transport coefficients, $t_p$'s have to be decided, as mentioned before, with equation (6) and (7). The solution is

$$\begin{cases} t_1 = \frac{1-t_0}{5} \\ t_2 = \frac{1-t_0}{20} \end{cases}. \quad (35)$$

From these expressions, we first give the formulas for the calculation of equilibrium distribution of different particles and the products of the particle reservoir as follows,

$$N_{0i}^{(0)} = t_0 \rho \,, \quad (36)$$

$$N_{1i}^{(0)} = m_0 + d_1 + \rho[\frac{1}{3}(u_\alpha c_{1i\alpha}) + \frac{1}{2}(u_\alpha c_{1i\alpha})^2 - \frac{1}{6}u^2] \,, \quad (37)$$

$$N_{2i}^{(0)} = m_0 + d_2 + \rho[\frac{1}{12}(u_\alpha c_{2i\alpha}) + \frac{1}{8}(u_\alpha c_{2i\alpha})^2 - \frac{1}{24}u^2] \,, \quad (38)$$

$$R = -8m_0 - \frac{2}{3}\rho u^2 \,. \quad (39)$$

The average density of the speed-1 particle is kept to be 4 times of that of the speed-2 particle so as to result in an isotropic fourth order tensor. Referring to those macroscopic transport coefficients, they can also be written explicitly as functions of $t_0$ and $\tau$,

$$c_s = \sqrt{\frac{3}{5}(1 - t_0)} \,, \quad (40)$$

$$\eta = \frac{1}{3}(\tau - \frac{1}{2}) \,, \quad (41)$$

$$\zeta = (\tau - \frac{1}{2})(\frac{2}{3} - c_s^2) \,. \quad (42)$$

The sound speed can be tuned to a suitable value by properly selecting the $t_0$. We observed that the dependence on $(1 - t_0)$ borne by $c_s$ in equation (40) is physically meaningful. When $t_0 = 0$, or there are no rest particles on the lattice field, the sound speed would reach its maximum value. On the contrary, as $t_0 = 1$, the expression indicates a zero sound speed. Because in the latter case, variations in the density would be absorbed completely into the rest particles so that no sound wave could be transported. The bulk viscosity's dependence on $c_s$ is nontrivial, as indicated in the analysis of the standard fluid mechanics(Landu and Lifshitz, 1959). Finally the positivity of the shear viscosity requires that $\tau$ should always be larger than $\frac{1}{2}$.

# 4 NUMERICAL EXPERIMENTS

## 4.1 Measurement Of Transport Coefficients

The method used to measure the transport coefficient of the modeled fluid is described as follows. First a periodically perturbative field of mass and momentum should be set up artificially, then the relaxing process would be measured and spatially Fourier-transformed. With the use of the curve fit technique, the relation between the wave vector $\boldsymbol{k}$ and the angular frequency $\omega$ could be found out, from which the values of various transport coefficients can be generalized.

For simplicity, the wave vector $\boldsymbol{k}$ is set to be $(k, 0)$ while the uniform motion of the modeled fluid is characterized by $(u_0, 0)$. Thus the perturbated density and momentum can be written as

$$\rho = \rho_0 + \delta\rho e^{i(\omega t + kx)} \,, \quad (43)$$

$$\rho u_x = \rho u_0 + \delta j_x e^{i(\omega t + kx)} \,, \quad (44)$$

$$\rho u_y = \delta j_y e^{i(\omega t + kx)} \,. \quad (45)$$

Here $\delta\rho$, $\delta j_x$ and $\delta j_y$ are small quantities which decide the amplitude of the perturbation. When these expressions are substituted into the macro-dynamic equations and higher order terms are dropped, a set of linearized hydrodynamic equations emerge as follows,

$$i\omega\, \delta\rho + ik\, \delta j_x = 0 \,, \quad (46)$$

$$i\omega\, \delta j_x + ik\,(u_0 \delta\rho + 2\delta j_x) g u_0 = -ik\, c_s^2 \delta\rho - k^2(\eta + \zeta)\delta j_x \,, \quad (47)$$

$$i\omega\, \delta j_y + ik\, g u_0 \delta j_y = -k^2 \eta \delta j_y \,. \quad (48)$$

We have deliberately left an additional coefficient before the advection term. If the Galilean invariance is recovered by the model, it should be 1. In the sequel, we shall use proper combination of equations (46), (47) and (48) to measure the sound speed $c_s$, the shear viscosity $\eta$, the bulk viscosity $\zeta$ and the advection coefficient $g$ of the modeled fluid.

### 4.1.1 Sound Speed

To measure the sound speed, we need to set $u_0 = 0$ and neglect the effect of viscosity. We use equations

(46) and (47) to obtain the following expression if only the density perturbation is considered,

$$c_s = \frac{\omega_\rho}{k} \,. \tag{49}$$

Here $\omega_\rho$ denote the angular frequency for the damping of the density perturbations. The numerical result is shown in figure 2. We see that the difference to the theoretical prediction is trivial.

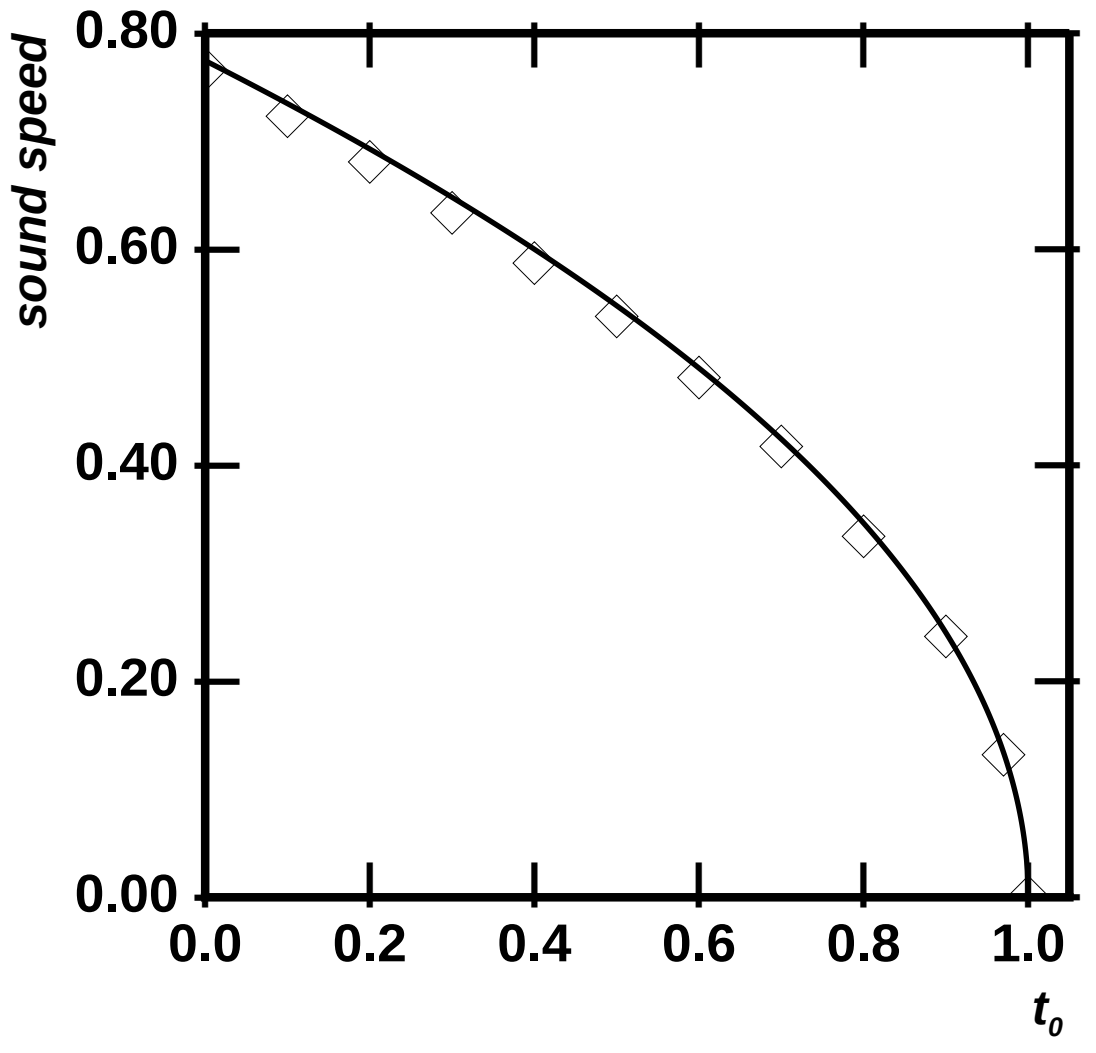

Figure 2: Sound speed of the modeled fluid as a function of $t_0$. The solid line is the theoretical prediction.

4.1.2 Shear and Bulk Viscosity

The expression of the shear viscosity can easily be obtained from equation (48) with $u_0 = 0$,

$$\eta = -\frac{i\omega_{j_y}}{k^2} \,. \tag{50}$$

Here $\omega_{j_y}$ refers to the angular frequency for the damping of the momentum perturbation in the $y$-direction. In the same spirit we can derive the formula for bulk viscosity from equation (46) and (47), by considering the damping of the momentum perturbation in the $x$-direction,

$$\zeta = \frac{\omega_{j_x}^2 - k^2 c_s^2}{i\omega_{j_x} k^2} - \eta \,. \tag{51}$$

Here we assume that the sound speed and the shear viscosity have already been known. The comparison between results of the numerical measurements and the theoretical predictions is shown in figure 3 and figure 4, respectively.

There obviously exists deviation from theoretical solution in the shear viscosity when the relaxation time becomes large enough. This is due to the large Knudsen number which is characterized by a slower approaching to the equilibrium state through the collision process. When the Knudsen number increases, the mean free path of particles would get to be comparable with the system size so that the hydrodynamic mode would be broken down eventually.

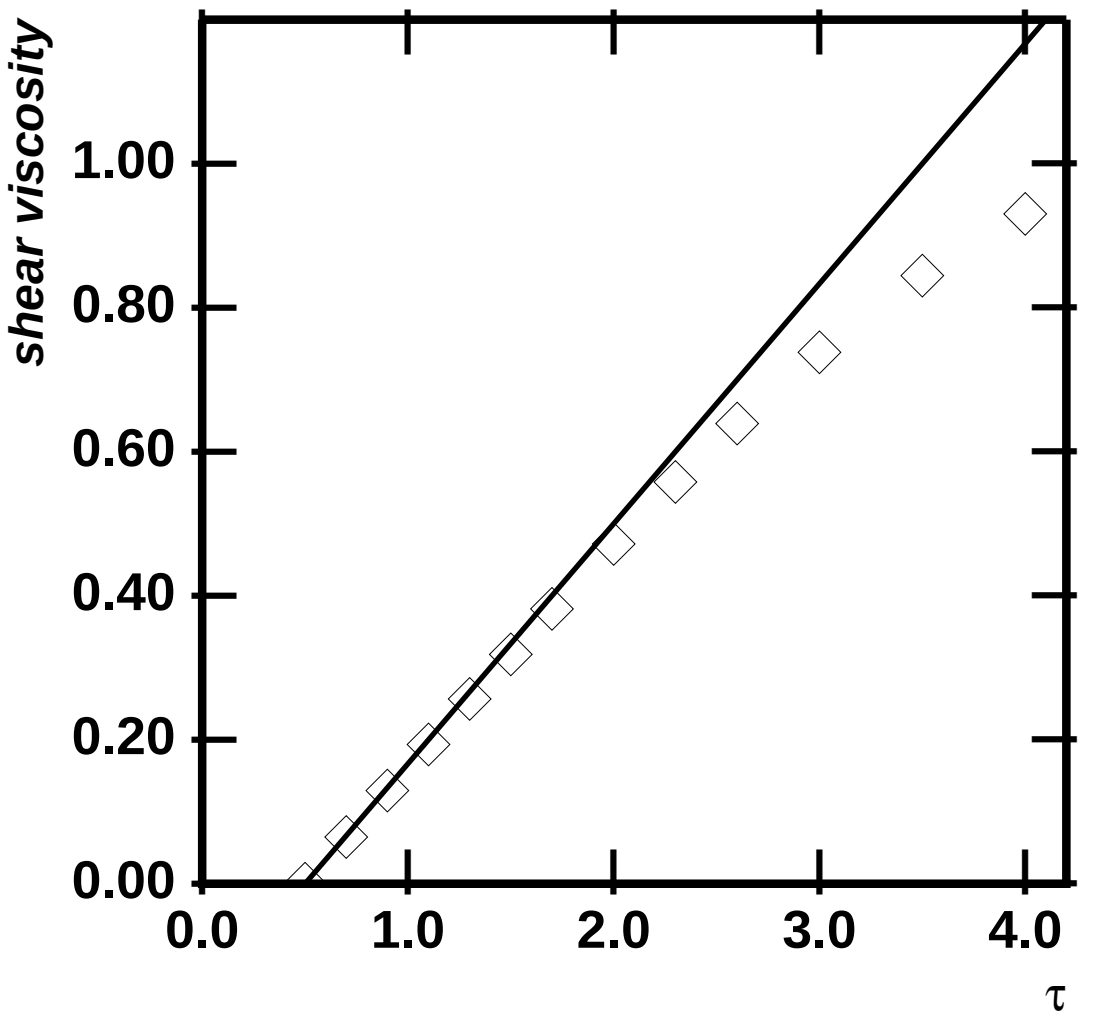

Figure 3: Shear viscosity of the modeled fluid as a function of $\tau$. The solid line is the theoretical prediction.

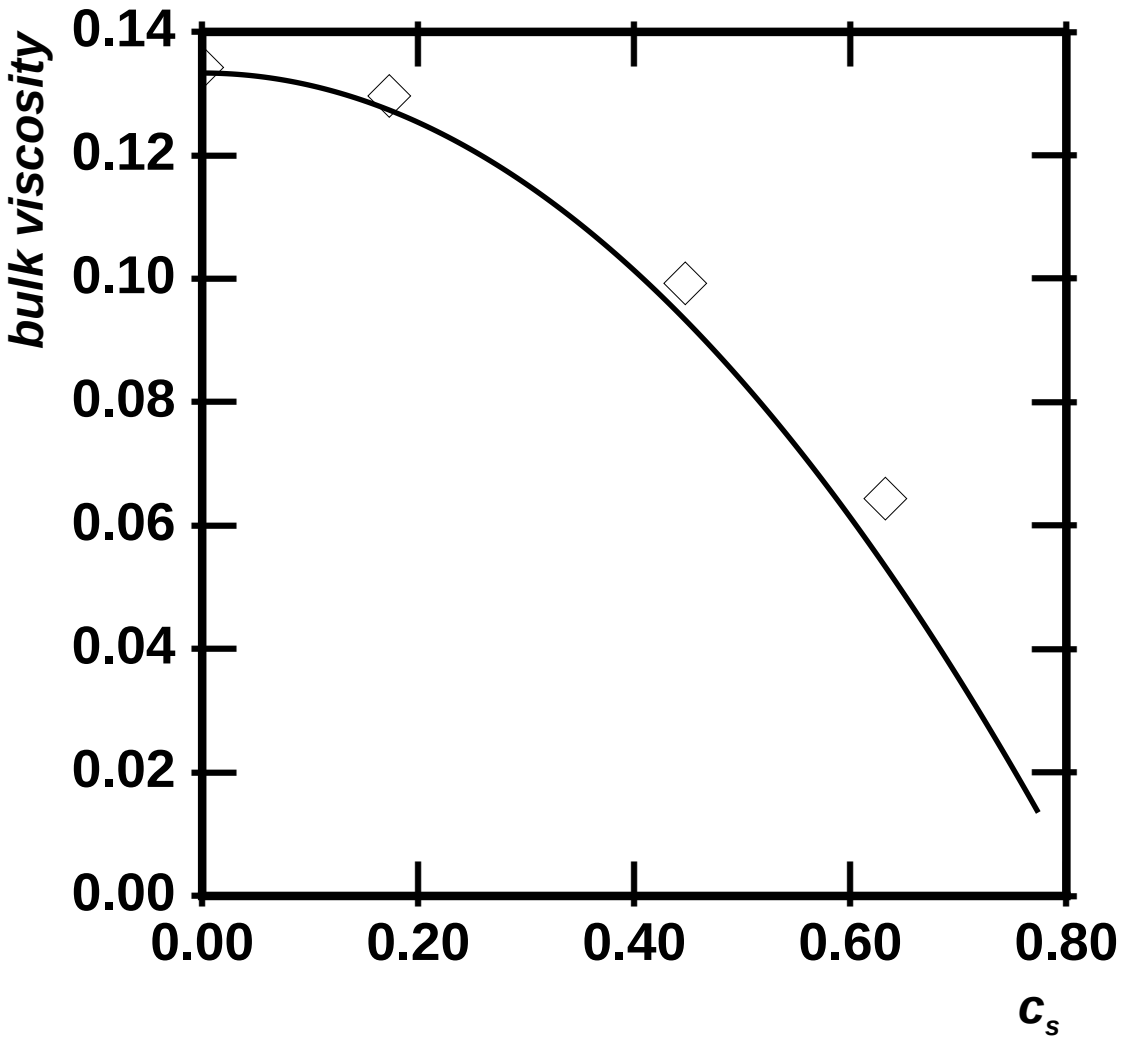

Figure 4: Bulk viscosity of the modeled fluid as a function of sound speed. The solid line is the theoretical prediction.

#### 4.1.3 Advection Coefficient

We initialize the flow motion with a nonvanishing $u_0$, and switch off the viscous effects in the mean time. Perturbations are added to the momentum density in the $y$-direction. From equation (48) we obtain,

$$g = -\frac{w_{j_y}}{k u_0} \tag{52}$$

The result is printed in figure 5, which can be taken as a complete proof of the recovery of the Galilean invariance in the modeled fluid.

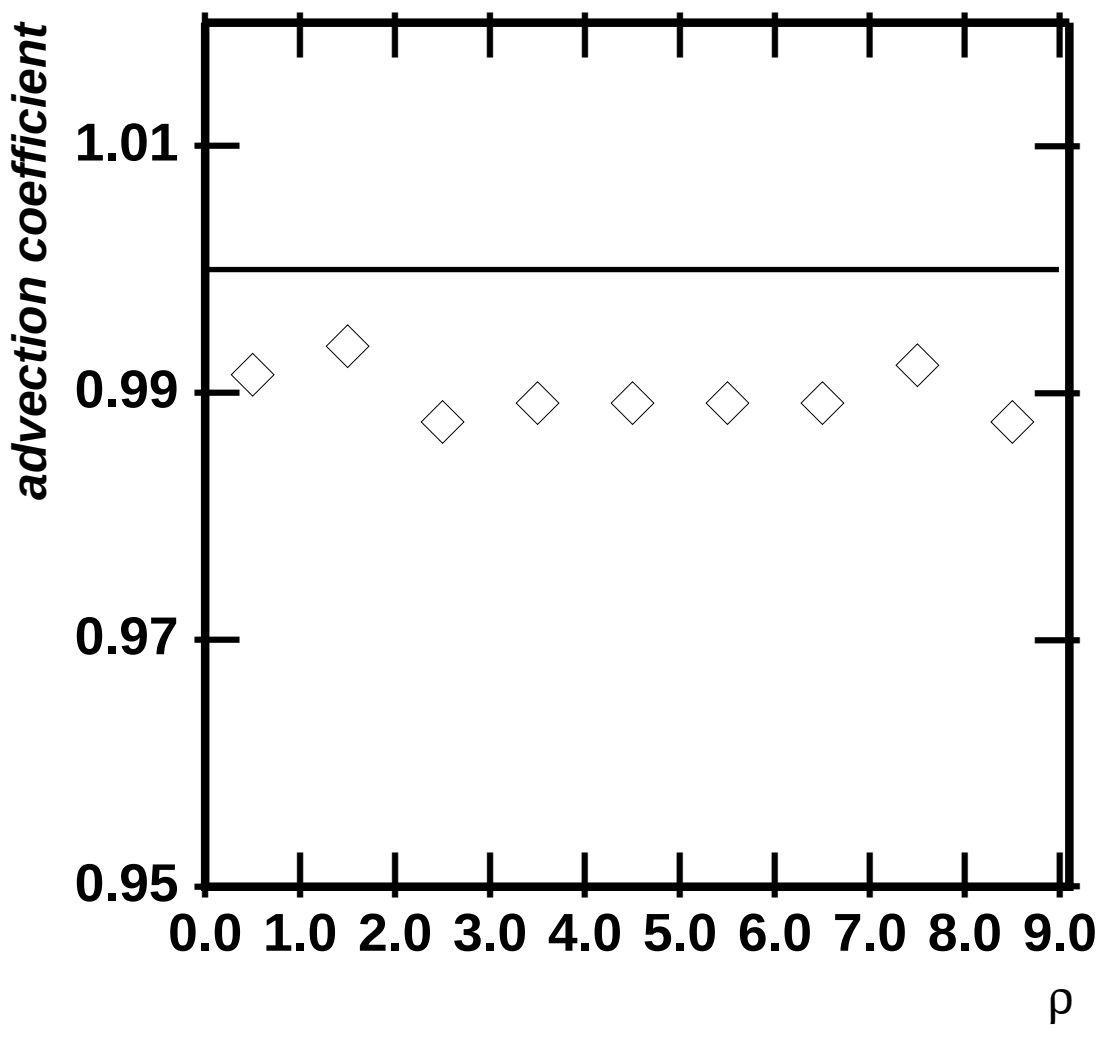

Figure 5: Advection coefficient of the modeled fluid as a function of $\rho$. The solid line is the theoretical prediction.

## 5 CONCLUSION

We extended the range of application of a multi-speed LBE model by introducing one degree of freedom in the sound speed of the modeled fluid. The analytical results seem to be physically meaningful while the numerical verification show good agreement with them.

When the sound speed of the modeled fluid is tuned to be comparable with the flow velocity, the Mach number could be much higer than the conventional scheme. This would make it easier for the model to simulate the supersonic flows. Moreover, the controller of the sound speed, namely $t_0$, need not to be constant. We may design it to be properly dependent on thermodynamic variables(Chen *et al.*, 1993) so that some realistic isentropic flow problems can be simulated either.

On the other hand, because the temperature has not been defined in the model, many practical applications are not amenable so far. We think that the difficulty would be overcome in the near future.